\documentclass[aps,epsf,showpacs,usenatbib]{iopart}
\pdfoutput=1
\usepackage{graphics}
\usepackage{graphicx}
\usepackage{epsfig}
\usepackage{times}
\usepackage{subfigure}
\usepackage{psfrag}
\usepackage{amssymb}

\def\reff@jnl#1{{\rm#1\/}}
\def\aj{\reff@jnl{AJ}}                 
\def\araa{\reff@jnl{ARA\&A}}           
\def\apj{\reff@jnl{ApJ}}               
\def\apjl{\reff@jnl{ApJ}}              
\def\apjs{\reff@jnl{ApJS}}             
\def\ao{\reff@jnl{Appl.Optics}}        
\def\apss{\reff@jnl{Ap\&SS}}           
\def\aap{\reff@jnl{A\&A}}              
\def\aapr{\reff@jnl{A\&A~Rev.}}        
\def\aaps{\reff@jnl{A\&AS}}            
\def\azh{\reff@jnl{AZh}}               
\def\baas{\reff@jnl{BAAS}}             
\def\jrasc{\reff@jnl{JRASC}}           
\def\memras{\reff@jnl{MmRAS}}          
\def\mnras{\reff@jnl{MNRAS}}           
\def\pra{\reff@jnl{Phys.Rev.A}}        
\def\prb{\reff@jnl{Phys.Rev.B}}        
\def\prc{\reff@jnl{Phys.Rev.C}}        
\def\prd{\reff@jnl{Phys.Rev.D}}        
\def\prl{\reff@jnl{Phys.Rev.Lett}}     
\def\pasp{\reff@jnl{PASP}}             
\def\pasj{\reff@jnl{PASJ}}             
\def\qjras{\reff@jnl{QJRAS}}           
\def\skytel{\reff@jnl{S\&T}}           
\def\solphys{\reff@jnl{Solar~Phys.}}   
\def\sovast{\reff@jnl{Soviet~Ast.}}    
\def\ssr{\reff@jnl{Space~Sci.Rev.}}    
\def\zap{\reff@jnl{ZAp}}               
\def\nat{\reff@jnl{Nature}}            
\def\cqg{\reff@jnl{Class. Quantum Grav.}}    

\begin{document}
\input epsf.tex

\title[{\sc MultiNest} for gravitational wave data analysis]{Use of the {\sc MultiNest} algorithm for gravitational wave data analysis}
\author{Farhan Feroz$^{1}$, Jonathan R. Gair$^{2}$, Michael P. Hobson$^{1}$ and Edward K. Porter$^{3}$}
\address{$^{1}$Astrophysics Group, Cavendish Laboratory, JJ Thomson Avenue, Cambridge CB3 0HE, UK}
\address{$^{2}$Institute of Astronomy, Madingley Road, Cambridge CB3 0HA, UK} 
\address{$^{3}$APC, UMR 7164, Universit\'{e} Paris 7 Denis Diderot, 10, rue Alice Domon et L\'{e}onie Duquet, 75205 Paris Cedex 13, France}

\vspace{1cm}
\begin{abstract}
We describe an application of the {\sc MultiNest} algorithm to gravitational wave data analysis. {\sc MultiNest} is a multimodal nested sampling algorithm designed to efficiently evaluate the Bayesian
evidence and return posterior probability densities for likelihood surfaces containing multiple secondary modes. The algorithm employs a set of `live' points which are updated by partitioning the set into
multiple overlapping ellipsoids and sampling uniformly from within them. This set of `live' points climbs up the likelihood surface through nested iso-likelihood contours and the evidence and posterior
distributions can be recovered from the point set evolution. The 
algorithm is model-independent in the sense that the specific problem being tackled enters only through the likelihood computation, and does
not change how the `live' point set is updated. In this paper, we consider the use of the algorithm for gravitational wave data analysis by searching a simulated LISA data set containing two non-spinning supermassive black hole binary signals. The algorithm is able to rapidly identify all the modes of the solution and recover the true parameters of the sources to high precision.
\end{abstract}
\pacs{04.25.Nx, 04.30.Db, 04.80.Cc}

\maketitle

\section{Introduction}
There is currently much active research into data analysis algorithms for gravitational wave detectors, both on the ground and for the proposed space-based gravitational wave detector, the Laser
Interferometer Space Antenna (LISA)~\cite{lisa}. Research into LISA data analysis is being encouraged by the Mock LISA Data Challenges~\cite{mldc} (MLDC), which so far have included data sets containing
individual and multiple white-dwarf binaries, a realisation of the whole galaxy of compact binaries, single and multiple non-spinning supermassive black hole (SMBH) binaries, either isolated or on top of a
galactic confusion background and isolated extreme-mass-ratio inspiral (EMRI) sources in purely instrumental noise. While most of these Challenges have been solved in the sense that several groups returned
solutions that matched the injected parameters, the EMRI case caused particular difficulty~\cite{EtfAG,neilemri,BBGPa,BBGPb}. These problems arose primarily because the likelihood surface contains many
secondary maxima, and so algorithms such as Markov Chain Monte Carlo (MCMC) tended to become stuck on secondary maxima and were unable to find the primary mode. In Challenge 3 of the MLDC, two new sources were
introduced that have similar multi-modal likelihood surfaces --- spinning black hole binaries and cosmic string cusps~\cite{key}. The final analysis of the LISA data will also have to contend with the
presence of multiple overlapping sources simultaneously present in the data stream. For these reasons, it is desirable to have algorithms that are able to simultaneously identify and characterize all the
modes of the likelihood surface. One approach is to use an evolutionary algorithm, and a recent implementation of these ideas is described in~\cite{hea}. However, there are also existing tools that have been
developed for other applications that have the necessary features to be useful for the gravitational wave case. One such algorithm is {\sc MultiNest}~\cite{feroz08,multinest}, which is a multi-modal nested
sampling algorithm.

The nested sampling algorithm~\cite{Skilling04} was developed as a tool for evaluating the Bayesian evidence. It employs a set of `live' points, each of which represents a particular set of parameters in the multi-dimensional search space. These points move as the algorithm progresses, and climb together through nested contours of increasing
likelihood.  At each step the algorithm works by finding a point of higher likelihood than the lowest likelihood point in the `live' point set and then replacing the lowest likelihood point with the new point.
Nested sampling has been applied to the issue of model selection for ground based observations of gravitational waves~\cite{Veitch:2008wd} and forms part of the evolutionary algorithm for LISA data analysis
discussed in~\cite{hea}. However, the primary difficulty in using nested sampling is to efficiently sample points of higher likelihood to allow the points to climb up the likelihood surface. 

{\sc MultiNest} solves the problem of efficient sampling by using the current set of `live' points as a model of the shape of the likelihood surface. The idea is to decompose the set of `live' points into a
sequence of overlapping ellipsoids, and then sample from within one of these ellipsoids which is chosen at random from the current set. The algorithm is model-free in the sense that the evolution of the `live'
point set does not make use of any knowledge about the properties of the likelihood surface. The only model-specific element is the subroutine which computes the likelihood at a given point in parameter
space. This approach reduces much of the computational overhead associated with evaluating Fisher Matrices etc. in order to move the `live' points on the likelihood surface. {\sc MultiNest} has proven to be
able to efficiently find and separate modes of a likelihood surface in many different applications~\cite{2008arXiv08100781F,2008arXiv08111199F,Feroz:2008wr,2008JHEP12024T}. This suggests it may be a powerful
tool for gravitational wave data analysis as well, and in this paper we explore these possibilities using a search for non-spinning SMBH binaries as a test case. Such sources are known to possess a degeneracy in the likelihood, since at low frequency the response of the detector to the true sky position and the point antipodal to it on the sky can not be distinguished. We can therefore use these sources as a test case to investigate the ability of {\sc MultiNest} to explore a multi-modal likelihood in the context of gravitational wave detection.

The coalescences of supermassive black holes (SMBHs) in binaries are expected to be a major source of gravitational waves (GW) for LISA, and we should be able to detect such systems out to a redshift of
$z \sim 10$ with intrinsic parameter errors of a fraction of a percent to a few percent and extrinsic parameter errors of a few to a few tens of percent~\cite{hugheslang, cornishporter1}. The detection problem for non-spinning SMBH binaries has already been solved, in the sense that several groups have been able
to successfully detect and recover parameters for such systems both in controlled studies~\cite{cornishporter1, cornishporter2} and in blind tests in the context of the MLDC. Several algorithms have been
proven to be successful --- Metropolis Hastings Monte Carlo (MHMC)~\cite{cornishporter1, cornishporter2}, template-bank based searches~\cite{babakstoch,harry}, time-frequency methods~\cite{mldcrd2} and
hierarchical searches mixing time-frequency and MHMC techniques~\cite{jplsmbh}. More recently, the evolutionary algorithm has also been shown to be effective in searches for non-spinning SMBH
binaries~\cite{hea}. Thus, although the application of {\sc MultiNest} to these systems will be an interesting test case, it is not essential to the success of LISA. However, the real power of the algorithm
will be in its application to multi-modal likelihood surfaces for systems such as EMRIs. The present work is an indication of the power of the technique and a stimulus for further research on this algorithm
in the context of gravitational wave data analysis. In addition, the evidence values returned by {\sc MultiNest} can be used for model selection, which will also be important for LISA. Bayesian model selection using MCMC techniques has already been explored in the LISA context~\cite{littenberg09} for the case of white-dwarf binaries. {\sc MultiNest} provides an alternative approach to addressing similar questions, although we will not discuss that application of the algorithm in the present paper.

The paper is organised as follows. In section~\ref{sec:bayesian} we give a brief introduction to Bayesian inference and then give details of the {\sc MultiNest} algorithm in section~\ref{sec:multinest}, including
an overview of nested sampling and a description of the ellipsoidal sampling scheme. In section~\ref{sec:wave} we briefly describe the waveform and noise models used in this analysis. In
section~\ref{sec:search} we describe an application of the algorithm to a simultaneous search for two non-spinning black hole binaries in a single data set. We provide results on the parameter recovery achieved by the
algorithm and compare these to the one sigma error estimates from the Fisher Matrix. We finish in section~\ref{sec:discuss} with a summary and a discussion of possible future applications of this work.

\section{Bayesian Inference}\label{sec:bayesian}

Our detection methodology is built upon the principles of Bayesian inference, and so we begin by giving a brief summary of this framework. Bayesian inference methods provide a consistent approach to the
estimation of a set of parameters~$\mathbf{\Theta}$ in a model (or hypothesis) $H$ for the data $\mathbf{D}$. Bayes' theorem states that
\begin{equation} \Pr(\mathbf{\Theta}|\mathbf{D}, H) =
\frac{\Pr(\mathbf{D}|\,\mathbf{\Theta},H)\Pr(\mathbf{\Theta}|H)}{\Pr(\mathbf{D}|H)},
\label{eq:bayes}
\end{equation}
where $\Pr(\mathbf{\Theta}|\mathbf{D}, H) \equiv P(\mathbf{\Theta})$ is the posterior probability distribution of the parameters, $\Pr(\mathbf{D}|\mathbf{\Theta}, H) \equiv \mathcal{L}(\mathbf{\Theta})$ is
the likelihood, $\Pr(\mathbf{\Theta}|H) \equiv \pi(\mathbf{\Theta})$ is the prior distribution, and $\Pr(\mathbf{D}|H) \equiv \mathcal{Z}$ is the Bayesian evidence.

Bayesian evidence is simply the factor required to normalise the posterior over~$\mathbf{\Theta}$ and is given by:
\begin{equation}
\mathcal{Z} =
\int{\mathcal{L}(\mathbf{\Theta})\pi(\mathbf{\Theta})}d^N\mathbf{\Theta},
\label{eq:Z}
\end{equation}
where $N$ is the dimensionality of the parameter space. Since the Bayesian evidence is independent of the parameter values~$\mathbf{\Theta}$, it is usually ignored in parameter estimation problems and the
posterior inferences are obtained by exploring the un--normalized posterior using standard MCMC sampling methods.

Bayesian parameter estimation has been used quite extensively in a variety of astronomical applications, including gravitational wave astronomy, although standard MCMC methods, such as the basic Metropolis--Hastings algorithm or the Hamiltonian sampling technique (see e.g. \cite{MacKay}), can experience problems in sampling efficiently from a multi--modal posterior distribution or one with large (curving) degeneracies between parameters. Moreover, MCMC methods often require careful tuning of the proposal distribution to sample efficiently, and testing for convergence can be problematic.

In order to select between two models $H_{0}$ and $H_{1}$ one needs to compare their respective posterior probabilities given the observed data set $\mathbf{D}$, as follows:
\begin{equation}
\frac{\Pr(H_{1}|\mathbf{D})}{\Pr(H_{0}|\mathbf{D})}
=\frac{\Pr(\mathbf{D}|H_{1})\Pr(H_{1})}{\Pr(\mathbf{D}|H_{0})\Pr(H_{0})}
=\frac{\mathcal{Z}_1}{\mathcal{Z}_0}\frac{\Pr(H_{1})}{\Pr(H_{0})},
\label{eq:model_select}
\end{equation}
where $\Pr(H_{1})/\Pr(H_{0})$ is the prior probability ratio for the two models, which can often be set to unity but occasionally requires further consideration (see, for example,
\cite{2008arXiv08100781F,2008arXiv08111199F} for the cases where the prior probability ratios should not be set to unity). It can be seen from Eq.~(\ref{eq:model_select}) that the Bayesian evidence plays a
central role in Bayesian model selection. As the average of likelihood over the prior, the evidence automatically implements Occam's razor: a simpler theory which agrees well enough with the empirical
evidence is preferred. A more complicated theory will only have a higher evidence if it is significantly better at explaining the data than a simpler theory.

Unfortunately, evaluation of Bayesian evidence involves the multidimensional integral (Eq.~(\ref{eq:Z})) and thus presents a challenging numerical task. Standard techniques like thermodynamic integration
\cite{Ruanaidh,littenberg09} are extremely computationally expensive which makes evidence evaluation typically at least an order of magnitude more costly than parameter estimation. Some fast approximate methods have been
used for evidence evaluation, such as treating the posterior as a multivariate Gaussian centred at its peak (see, for example, \cite{Hobson02}), but this approximation is clearly a poor one for highly non-Gaussian and
multi--modal posteriors. Various alternative information criteria for model selection are discussed in \cite{Liddle07}, but the evidence remains the preferred method.

\section{Nested Sampling and the {\sc MultiNest} Algorithm}\label{sec:multinest}

%
\begin{figure}
\begin{center}
\subfigure[]{\includegraphics[width=0.3\columnwidth]{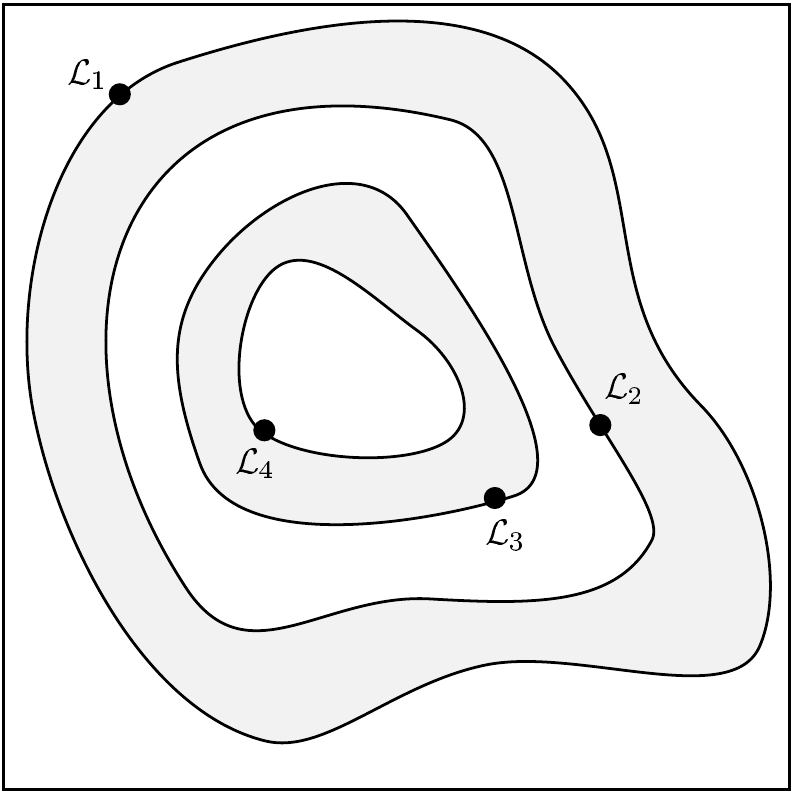}}\hspace{0.5cm}
\subfigure[]{\includegraphics[width=0.3\columnwidth]{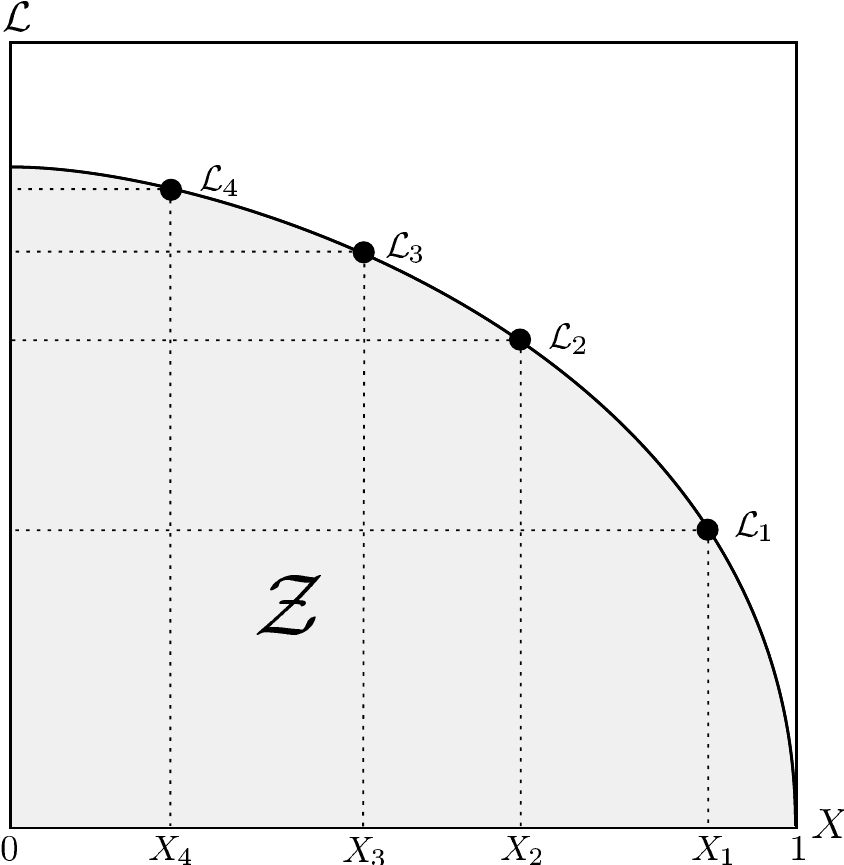}}
\caption{Cartoon illustrating (a) the posterior of a two dimensional problem; and (b) the transformed $\mathcal{L}(X)$  function where the prior volumes $X_{i}$ are associated with each likelihood
$\mathcal{L}_{i}$.}
\label{fig:NS}
\end{center}
\end{figure}

Nested sampling~\cite{Skilling04} is a Monte Carlo method targetted at the efficient calculation of the evidence, but also produces posterior inferences as a by-product. It calculates the evidence by
transforming the multi--dimensional evidence integral into a one--dimensional integral that is easy to evaluate numerically. This is accomplished by defining the prior volume $X$ as $dX =
\pi(\mathbf{\Theta})d^D \mathbf{\Theta}$, so that
\begin{equation}
X(\lambda) = \int_{\mathcal{L}\left(\mathbf{\Theta}\right) > \lambda} \pi(\mathbf{\Theta}) d^N\mathbf{\Theta},
\label{eq:Xdef}
\end{equation}
where the integral extends over the region(s) of parameter space contained within the iso-likelihood contour $\mathcal{L}(\mathbf{\Theta}) = \lambda$. The evidence integral, Eq.~(\ref{eq:Z}), can then be
written as
\begin{equation}
\mathcal{Z}=\int_{0}^{1}{\mathcal{L}(X)}dX,
\label{eq:nested}
\end{equation}
where $\mathcal{L}(X)$, the inverse of Eq.~(\ref{eq:Xdef}), is a  monotonically decreasing function of $X$.  Thus, if one can evaluate the likelihoods $\mathcal{L}_{i}=\mathcal{L}(X_{i})$, where $X_{i}$ is a
sequence of decreasing values,
\begin{equation}
0<X_{M}<\cdots <X_{2}<X_{1}< X_{0}=1,
\end{equation}
as shown schematically in Fig.~\ref{fig:NS}, the evidence can be approximated numerically using standard quadrature methods as a weighted sum
\begin{equation}
\mathcal{Z}={\textstyle {\displaystyle \sum_{i=1}^{M}}\mathcal{L}_{i}w_{i}},
\label{eq:NS_sum}
\end{equation}
where the weights $w_{i}$ for the simple trapezium rule are given by $w_{i}=\frac{1}{2}(X_{i-1}-X_{i+1})$. An example of a posterior in two dimensions and its associated function $\mathcal{L}(X)$ is shown in
Fig.~\ref{fig:NS}.

\subsection{Evidence Evaluation}\label{app:nested:evidence}

In order to evaluate the evidence value given in (Eq.~\ref{eq:NS_sum}), a set of $N$ `live' points are drawn
uniformly from the full prior $\pi(\mathbf{\Theta})$ with the initial prior volume $X_{0}$ set to $1$. The
likelihood value for each of the $N$ `live' points is calculated and the point with the minimum likelihood value
$\mathcal{L}_{0}$ is removed from the `live' point set and is replaced by another point sampled uniformly from
the prior with likelihood $\mathcal{L} > \mathcal{L}_{0}$. This results in the reduction in the prior volume
within the iso-likelihood contour by a factor $t_{1}$ i.e. $X_{1} = t_{1} X_{0}$ where $t_{1}$ follows the
distribution of the largest of the $N$ samples drawn uniformly from the interval $[0,1]$ which is given by
$\Pr(t) = Nt^{N-1}$. This procedure is repeated at each subsequent iteration $i$, at which the point with the
lowest likelihood value $\mathcal{L}_{i}$ is replaced by a new point sampled uniformly from the prior with
$\mathcal{L} > \mathcal{L}_{i}$ until the entire prior volume has been traversed. The expected value and the standard
deviation of $\log t$, which dominates the geometrical exploration is given by: 
\begin{equation}
E[\log t] = -1/N, \quad \sigma[\log t] = 1/N.
\end{equation}
With the values of $\log t$ at different iterations being independent of each other, the prior volume at
iteration $i$ is expected to be:
\begin{equation}
\log X_{i} \approx -(i\pm\sqrt{i})/N.
\end{equation}
Thus, one takes
\begin{equation}
X_{i} = \exp(-i/N).
\label{eq:NS_prior}
\end{equation}
%

\subsection{Stopping Criterion}\label{nested:stopping}

The primary aim of nested sampling algorithm is to calculate the evidence value and so it should be terminated
when the evidence value has been calculated to required accuracy. One way to achieve this is by checking whether
the change in the evidence value from one iteration to the other is smaller than a specified tolerance but in
cases where the posterior contains narrow peaks close to its maximum, this can result in an underestimated
evidence value.

Skilling \cite{Skilling04} provides a robust stopping condition by calculating an upper limit on the evidence
that can be determined from the set of current `live' points. This limit is calculated by assuming that all the
`live' points have the likelihood value equal to the maximum--likelihood $\mathcal{L}_{\rm max}$ and so the
largest evidence contribution that can be made by the remaining portion of the posterior is
$\Delta{\mathcal{Z}}_{\rm i} = \mathcal{L}_{\rm max} X_{\rm i}$. $\mathcal{L}_{\rm max}$ can be taken to be the
maximum likelihood value amongst the `live' points. At the beginning of the nested sampling algorithm, it is
possible for the high likelihood regions of the posterior to not have been explored adequately resulting in the
maximum--likelihood value $\mathcal{L}_{\rm max}$, amongst the `live' points being a lot smaller than the true
maximum--likelihood value. The remaining prior volume $X_{\rm i}$ on the other hand would be quite high and
therefore the largest evidence contribution $\Delta{\mathcal{Z}}_{\rm i} = \mathcal{L}_{\rm max}X_{\rm i}$ would
be high as well allowing the algorithm to proceed. At subsequent stages, the remaining prior volume $X_{\rm i}$
decreases exponentially as given by (Eq.~\ref{eq:NS_prior}) and the high likelihood regions of the posteriors are
explored in greater detail with estimated $\mathcal{L}_{\rm max}$ getting closer and closer to the true
maximum--likelihood value. The largest evidence contribution $\Delta{\mathcal{Z}}_{\rm i} = \mathcal{L}_{\rm max}
X_{\rm i}$, thus provides a robust stopping criteria.

We choose to stop when this  quantity would no longer change the final evidence estimate by 0.5 in log--evidence.

\subsection{Posterior Inferences}\label{nested:posterior}

Although the nested sampling algorithm has been designed to calculate the evidence value, accurate posterior
inferences can also be generated as a by-product. Once the evidence $\mathcal{Z}$ is found, each one of the final
`live' points and the discarded points from the nested sampling process, i.e., the points with the lowest
likelihood value at each iteration $i$ of the algorithm is simply assigned the probability weight 
\begin{equation}
p_{i}=\frac{\mathcal{L}_{i}w_{i}}{\mathcal{Z}}.
\label{eq:12}
\end{equation} 
These samples can then be used to calculate inferences of posterior parameters such as  means, standard
deviations, covariances and so on, or to construct marginalised posterior distributions.

\subsection{{\sc MultiNest} Algorithm}\label{sec:method:bayesian:multinest}

%
\begin{figure}
\begin{center}
\subfigure[]{\includegraphics[width=0.42\columnwidth,height=0.42\columnwidth]{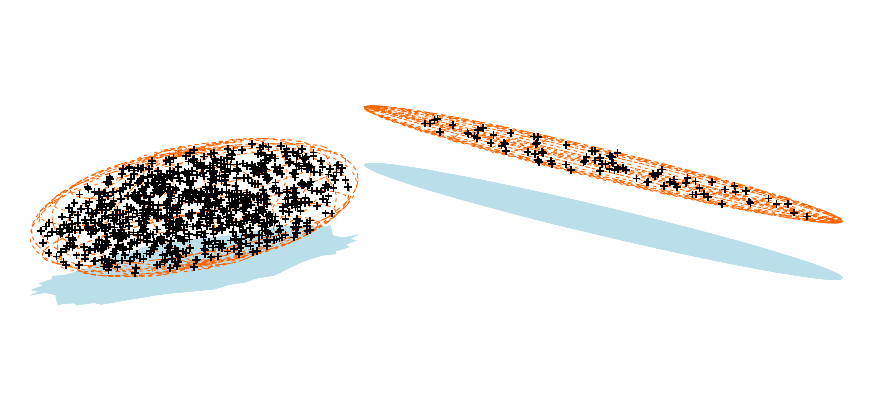}}\setcounter{subfigure}{1}
\subfigure[]{\includegraphics[width=0.42\columnwidth,height=0.42\columnwidth]{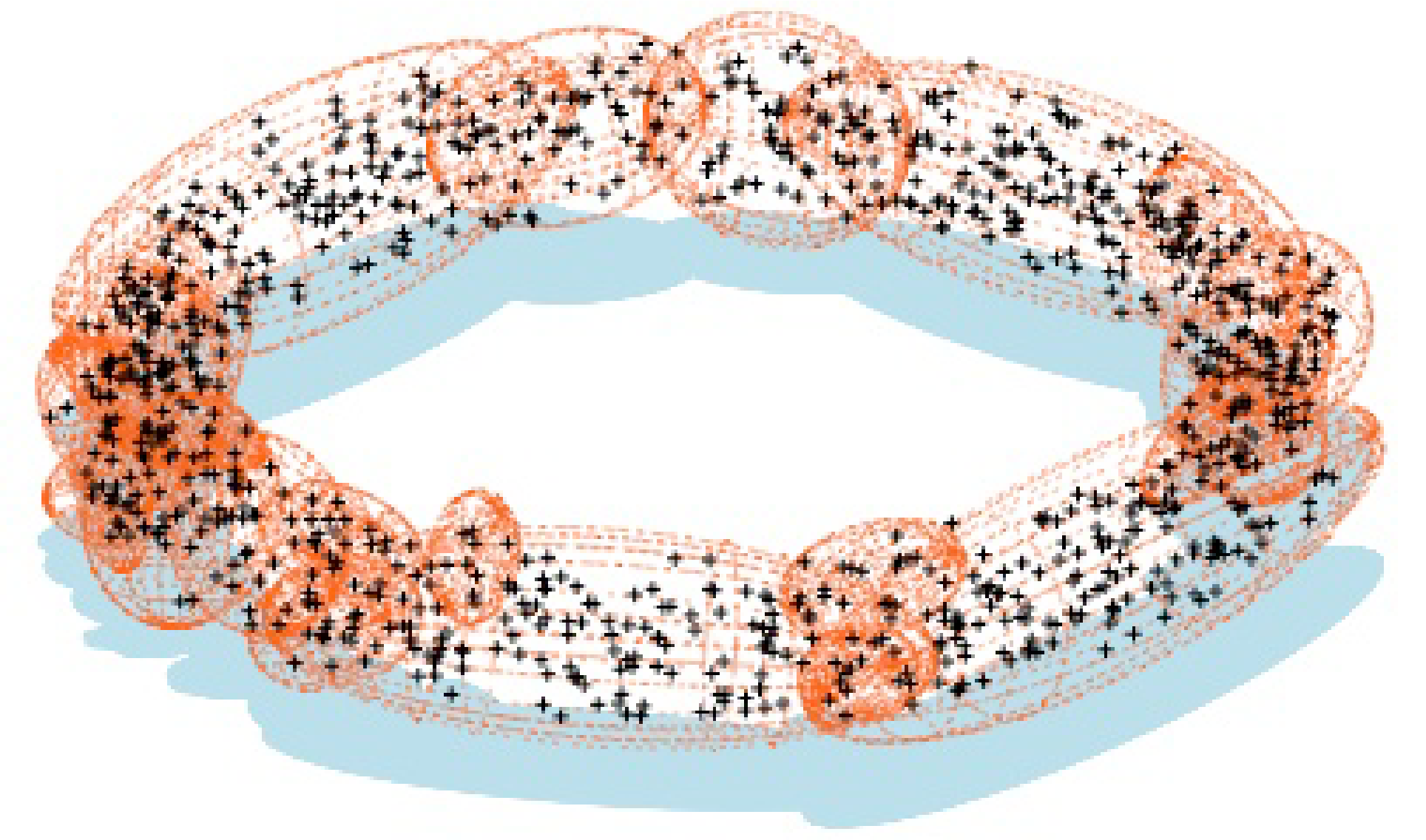}}\setcounter{subfigure}{2}
\caption{Illustrations of the ellipsoidal decompositions performed by {\sc MultiNest}. The points given as input are overlaid on the resulting ellipsoids. 1000 points were sampled uniformly from: (a) two
non-intersecting ellipsoids; and (b) a torus.}
\label{fig:dino}
\end{center}
\end{figure}
The most challenging task in implementing the nested sampling algorithm is drawing samples from the prior within the hard constraint $\mathcal{L}> \mathcal{L}_i$ at each iteration $i$. Employing a naive
approach that draws blindly from the prior would result in a steady decrease in the acceptance rate of new samples with decreasing prior volume (and increasing likelihood). The {\sc MultiNest} algorithm
\cite{feroz08,multinest} tackles this problem through an ellipsoidal rejection sampling scheme by enclosing the `live' point set within a set of (possibly overlapping)  ellipsoids and a new point is then drawn
uniformly from the region enclosed by these ellipsoids. The number of points in an individual ellipsoid and the total number of ellipsoids is  decided by an `expectation--maximization' algorithm so that the total
sampling volume, which is equal to the sum of volumes of the ellipsoids, is minimized. This allows maximum flexibility and efficiency by breaking up a mode resembling a Gaussian into a relatively small number of
ellipsoids, and if the posterior mode possesses a pronounced curving  degeneracy so that it more closely resembles a (multi--dimensional) `banana' then it is broken into  a relatively large number of small `overlapping' ellipsoids (see Fig.~\ref{fig:dino}). With enough `live' points, this approach allows the detection of all the modes simultaneously, resulting in typically two orders of magnitude improvement in
efficiency and accuracy over standard methods for inference problems in cosmology and particle physics phenomenology (see, for example, \cite{2008arXiv08100781F,2008arXiv08111199F,Feroz:2008wr,2008JHEP12024T}).

The ellipsoidal decomposition scheme described above also provides a mechanism for mode identification. By forming chains of overlapping ellipsoids (enclosing the `live' points), the algorithm can identify distinct
modes with distinct ellipsoidal chains, e.g., in Fig.~\ref{fig:dino} panel (a) the algorithm identifies two distinct modes while in panel (b) the algorithm identifies only one mode as all the ellipsoids are
linked with each other because of the overlap between them. Once distinct modes have been identifed, they are evolved independently.

Another feature of the {\sc MultiNest} algorithm is the evaluation of the global as well as the `local' evidence values associated with each mode. These evidence values can be used in calculating the
probability that an identified `local' peak in the posterior corresponds to a real object (see, for instance, \cite{Veitch:2008wd,2008arXiv08100781F,2008arXiv08111199F}). We defer the discussion of quantifying the SMBH detection to a later work.

\section{Problem Description}\label{sec:wave}
\subsection{Waveform model}
The waveform from a binary composed of two non-spinning black holes depends on nine parameters: $\vec{\lambda}=\{\ln(M_{c}),\ln(\mu),\theta, \phi, \ln(t_{c}), \iota, \varphi_{c}, \ln(D_{L}), \psi\}$, where
$M_{c}$ is the chirp mass, $\mu$ is the reduced-mass, $(\theta,\phi)$ are the sky location of the source, $t_{c}$ is the time-to-coalescence, $\iota$ is the inclination of the orbit of the binary,
$\varphi_{c}$ is the phase of the GW at coalescence, $D_{L}$ is the luminosity distance and $\psi$ is the polarization of the GW measured in the Solar System barycentre (SSB). $\psi$ is the angle between the principal polarisation axes of the source and the ``natural'' polarisation axes to use in the SSB, which are perpendicular to the line of sight to the source and the orbital angular momentum of the binary. We model the waveform using the
restricted post-Newtonian approximation, in which the two polarizations of the GW are~\cite{biww}
\begin{eqnarray}
h_{+}& =& \frac{2Gm\eta}{c^{2}D_{L}}\left(1+\cos^{2}\iota\right)\left(\frac{Gm\omega}{c^3}\right)^{\frac{2}{3}}\cos(\Phi),\\ \nonumber \\
h_{\times} &= &-\frac{4Gm\eta}{c^{2}D_{L}}\cos\iota\,\left(\frac{Gm\omega}{c^3}\right)^{\frac{2}{3}}\sin(\Phi),
\end{eqnarray}
where $m=m_{1}+m_{2}$ is the total mass of the binary, $\eta = m_{1}m_{2}/m^{2}$ is the reduced mass ratio, $G$ is Newton's constant and $c$ is the speed of light. The chirp mass and reduced mass are given in
terms of $m$ and $\eta$ as $M_{c}=m\eta^{3/5}$ and $\mu = m\eta$.  The function $\omega=d\Phi_{orb}/dt$ is the orbital frequency, and $\Phi =\varphi_{c}-\varphi(t) = 2\Phi_{orb}$ is the gravitational wave
phase. We take these to 2PN order
\begin{eqnarray}
\omega(t)&=&\frac{c^{3}}{8Gm}\left[\Theta^{-3/8}+\left(\frac{743}{2688}+\frac{11}{32}\eta\right)\Theta^{-5/8}-\frac{3\pi}{10}\Theta^{-3/4}\right.\nonumber\\
&+&\left.\left(\frac{1855099}{14450688}+\frac{56975}{258048}\eta+\frac{371}{2048}\eta^{2}\right)\Theta^{-7/8}\right]\label{eqn:freq}\\
\Phi(t) &=& \varphi_{c}-\frac{2}{\eta}\left[\Theta^{5/8}+\left(\frac{3715}{8064}+\frac{55}{96}\eta\right)\Theta^{3/8}-\frac{3\pi}{4}\Theta^{1/4}\right.\nonumber\\
&+&\left.\left(\frac{9275495}{14450688}+\frac{284875}{258048}\eta+\frac{1855}{2048}\eta^{2}\right)\Theta^{1/8}\right],\label{eqn:phase}
\end{eqnarray}
where
\begin{equation}
\Theta(t;t_{c}) = \frac{c^{3}\eta}{5Gm}\left(t_{c}-t\right).
\end{equation}
The above model describes the inspiral only and not the merger or ringdown phase. To avoid artifacts when taking Fourier transforms, we follow the approach introduced in~\cite{cornishporter1} and now used in the MLDC~\cite{mldc} and continue the inspiral until the orbit reaches the innermost stable circular orbit at $6M$, but employ a hyperbolic taper from $7M$ to ensure the waveform goes smoothly to zero as the end of the inspiral approaches. The taper has the value of $1$ for $r > 7M$ and then smoothly goes to zero at $r=6M$, where the waveform finishes. To implement the LISA response, we use the low-frequency approximation, as described in~\cite{cutlowfreq}. In the low-frequency limit we expect there to be bright modes in the likelihood at both the true sky position and at a position approximately antipodal to it, as mentioned earlier. The antipodal position corresponds to the shift $\theta \rightarrow \pi-\theta$, $\phi \rightarrow \phi+\pi$. The `antipodal' mode of the likelihood is strictly antipodal at low frequencies, but moves as the gravitational wave frequency increases.

The four parameters $\{D_{L}, \iota, \varphi_{c}, \psi\}$ are extrinsic parameters which only affect how the gravitational waveform phase at the detector projects into a detector response. It is possible to
search over these automatically using a generalisation of the $\cal{F}$-statistic~\cite{JKS}. Details of how the $\cal{F}$-statistic is computed for non-spinning SMBH binaries may be found in~\cite{cornishporter1}. We make use of the $\cal{F}$-statistic maximization in the first stage of our search.

A gravitational wave search is sensitive to redshifted masses, $\bar{m}_1 = (1+z)m_1$, rather than the intrinsic masses. All our subsequent results will be quoted for redshifted mass quantities, which we will denote with an overbar for clarity.

\subsection{Likelihood evaluation}
The gravitational waveform signals can be thought of as occupying a vector space, on which there is a natural scalar product~\cite{helst,owen}
\begin{equation}\label{eqn:scalarprod}
\left<h\left|s\right.\right> =2\int_{0}^{\infty}\frac{df}{S_{n}(f)}\,\left[ \tilde{h}(f)\tilde{s}^{*}(f) +  \tilde{h}^{*}(f)\tilde{s}(f) \right].
\label{eq:scalarprod}
\end{equation}
where 
\begin{equation}
\tilde{h}(f) = \int_{-\infty}^{\infty}\, dt\, h(t)e^{2\pi\imath ft}
\end{equation}
is the Fourier transform of the time domain waveform $h(t)$.  The quantity $S_{n}(f)$ is the one-sided noise spectral density of the detector. For a family of sources with waveforms~$h(t;{\vec{\lambda}})$
that depend on parameters~$\vec{\lambda}$, the output of the detector, $s(t) = h(t;\vec{\lambda_0}) + n(t)$, consists of the true signal~$h(t;\vec{\lambda_0})$ and a particular realisation of the noise,
$n(t)$. Assuming that the noise is stationary and Gaussian, the logarithm of the likelihood that the parameter values are given by~$\vec{\lambda}$ is
\begin{equation}
\log{\mathcal L}\left(\vec{\lambda}\right) = C-\left<s-h\left(\vec{\lambda}\right)\left|s-h\left(\vec{\lambda}\right)\right.\right>/2,  
\end{equation}
and it is this log-likelihood that is evaluated at each `live' point in the search. The constant, $C$, depends on the dimensionality of the search space, but its value is not important as we are only interested in the relative likelihoods of different points.

In section~\ref{sec:search} we will compare the errors in our recovered parameters to the theoretical noise-induced errors. The latter can be computed as $\sigma_i = \sqrt{(\Gamma^{-1})^{ii}}$ where
\begin{equation}
\Gamma_{ij} = \left<\frac{\partial h}{\partial \lambda^{i}}\left|\frac{\partial h}{\partial \lambda^{j}}\right.\right>
\end{equation}
is the Fisher Information Matrix (FIM).

\subsection{Noise model}
The noise consists of an instrumental part, which we take from~\cite{cornish}
\begin{eqnarray}
\fl S_{n}^{\rm inst}(f) &=&\frac{1}{4L^{2}}\left[ 2S_{n}^{\rm pos}(f) \left(2+\left(\frac{f}{f_{*}}^2\right)\right)\right. \nonumber\\
\fl &&+  \left.  8 S_{n}^{\rm acc}(f) \left(1+\cos^2\left(\frac{f}{f_{*}}\right)\right) \left( \frac{1}{(2\pi f)^{4}} + \frac{(2\pi 10^{-4} {\rm Hz})^{2}}{(2\pi f)^{6}}   \right)   \right] . 
\end{eqnarray}
where $L=5\times10^{6}$ km is the arm-length for LISA,  $S_{n}^{\rm pos}(f) = 4\times10^{-22}\,$m$^{2}/$Hz and $S_{n}^{\rm acc}(f) = 9\times10^{-30}\,$m$^{2}/$s$^{4}/$Hz are the position and acceleration noise
respectively. $S_{n}^{\rm inst}(f)$ has units of Hz$^{-1}$. The quantity $f_{*}=1/(2\pi L)$ is the mean transfer frequency for the LISA arms. In addition, there is a noise contribution from the confusion foreground of galactic compact binaries. We represent this using the model of~\cite{nyz,trc}
\begin{equation}
\fl S_{n}^{\rm conf}(f) = \left\{ \begin{array}{ll} 10^{-44.62}(f/{\rm Hz})^{-2.3} & 10^{-4} < (f/{\rm Hz}) \leq 10^{-3} \\ \\ 10^{-50.92}(f/{\rm Hz})^{-4.4} & 10^{-3} < (f/{\rm Hz})\leq 10^{-2.7}\\ \\ 10^{-62.8}(f/{\rm Hz})^{-8.8} &  10^{-2.7} < (f/{\rm Hz})\leq 10^{-2.4}\\ \\
10^{-89.68}(f/{\rm Hz})^{-20} &  10^{-2.4} < (f/{\rm Hz})\leq 10^{-2}  \end{array}\right.{\rm Hz}^{-1},
\end{equation}
The total noise is the sum of these contributions. At low frequencies, the LISA detector can be thought to consist of two independent right-angle
interferometers with independent noise described by the same power spectral density. The total likelihood is obtained by summing the scalar product~(\ref{eqn:scalarprod}) over the two detectors, and the FIM
of the network is similarly given by the sum of the two FIMs.

\section{Results of {\sc MultiNest} Search}
\label{sec:search}
To test the {\sc MultiNest} algorithm we generated a data set containing two SMBH binaries, one of which coalesced during the observation time, and one of which coalesced a few days after the end of the
observation. We took these sources to have the same parameters as the two sources searched for by the evolutionary algorithm in~\cite{hea} to facilitate direct comparison between the two techniques. The
parameters of the two sources are summarised in Table~\ref{tab:params}. We used a standard flat WMAP cosmology, with radiation, matter and dark energy densities $(\Omega_{R}, \Omega_{M}, \Omega_{\Lambda}) =
(4.9\times10^{-5}, 0.27, 0.73)$ and $H_{0}$=71 km/s/Mpc~\cite{wmap}. The redshifts of the sources therefore correspond to luminosity distances of 6.634Gpc and 5.024Gpc respectively. The redshifted chirp mass
and reduced mass are $(\bar{M_c},\bar{\mu}) = (4.9289\times10^6, 1.8182\times10^6)\,M_{\odot}$ for source 1, and $(2.997\times10^6, 1.44\times10^6)\,M_{\odot}$ for source 2, and the full, dual-TDI channel, signal-to-noise ratios (SNRs) are 200 and 131 respectively.

\begin{table}
\begin{center}
\lineup
\item[]\begin{tabular}{@{}*{10}{l}}
\br                              
$$&$m_1/M_{\odot}$&$m_2/M_{\odot}$&$t_{c}/yrs$&$\theta$&$\phi$&$z$&$\iota$&$\psi$&$\varphi_c$\cr 
\mr
1&$1\times10^7$ &$1\times10^6$&$0.90$&$0.6283$ &4.7124&1.0&1.1120&1.2330&2.2220\cr
2&$4\times10^6$&$1\times10^6$&$1.02$&$2.1206$&3.9429 & 0.8& 0.6565&
1.0646&1.5116\cr 
\br
\end{tabular}
\caption{Parameter values for the two SMBHBs considered in this study. The dual-TDI channel SNRs for the sources are 200 and 131 respectively.} 
\label{tab:params}
\end{center}
\end{table}

For the first stage of the search, we included the $\cal{F}$-statistic maximization in the likelihood evaluation and used {\sc MultiNest} with $1000$ `live' points to search only the five dimensional space of intrinsic parameters $(\bar{m_1}, \bar{m_2}, \theta, \phi, t_c)$. We used wide priors for this search of $\bar{m_1},\bar{m_2} \in \left[5\times10^5, 3\times 10^7 \right]M_{\odot}$, $t_c \in [0.85, 1.1]$yrs, $\theta \in \left[0,\pi\right]$ and $\phi \in \left[0,2\pi\right]$. In general, the number of `live' points required for a search is problem specific, depending on the dimensionality of the search space, and on the complexity of the likelihood surface. Therefore, the ideal number must be found by trial and error. Using too few `live' points leads to under-sampling of the posterior. It is normally clear from the posterior distributions whether the likelihood has been properly sampled or not. Using more `live' points than necessary does not affect the results, but is computationally more expensive. The choice of 1000 in this case was based on previous experience in other contexts, but appeared to work well. The algorithm returned 11 possible modes and, based on the time of coalesence, it was clear that five were associated with the coalescing source, and six with the non-coalescing source. Four modes were the same secondaries found using the evolutionary algorithm~\cite{hea}, each of which corresponded to mass and spin values close to those of one of the sources, but had sky position in the vicinity of one of the two sky solutions of the other source. Of the three additional secondaries identified in the likelihood, one was associated with the coalescing source, and was very close to the true values of the masses and $t_c$, but was also in the wrong position on the sky. The other two secondaries were associated with the non-coalescing source and had mass and $t_c$ parameters that were relatively close to the true values, although many Fisher Matrix $\sigma$ away, but had incorrect sky locations. It is likely that all of these secondaries correspond to parameter space points for which the waveforms match the brighter waveform cycles that come toward the end of the inspiral, but not the earlier part of the waveform. This stage of the search took approximately one and a half hours to run on a single 3.0 GHz Intel Woodcrest processor, and used $\sim 170,000$ likelihood evaluations.

In Table~\ref{tab:posterior_stage1} we list the mean and standard deviation of the posteriors recovered by the algorithm in this stage, for the two true and the two antipodal modes of the sources. In Table~\ref{tab:mnerrs} we assess the error in parameter recovery as a multiple of the theoretical error, computed from the Fisher matrix. This is one of the standard approaches used to assess entries to the LISA Mock Data Challenge and so we include it for easier comparison to other algorithms discussed in the literature. The posterior distributions are not Gaussian, which is assumed for the Fisher matrix estimate, so we do not necessarily expect these to be an accurate indication of the true error. In practice, we will estimate the uncertainty in the parameter estimation from the recovered posterior and, in all cases, the true solution lies within 1-$\sigma$ of the peak of the recovered posterior, when $\sigma$ is computed from the posterior in that way. We quote results in Table~\ref{tab:mnerrs} for the two true solutions, and for the two antipodal sky solutions. Errors are quoted for the redshifted chirp mass, $\bar{M_c}$, and reduced mass, $\bar{\mu}$, rather than $\bar{m_1}$ and $\bar{m_2}$, since the former are used more conventionally in the literature and this therefore facilitates comparison to other work, in particular the evolutionary algorithm~\cite{hea}.

We see that the algorithm recovered all of the intrinsic parameters to extremely high precision, being within $1\sigma$ of the true parameters for all but the antipodal sky solution of the second source. As mentioned earlier, the `antipodal' solution is no longer exactly antipodal on the sky at higher frequencies. Thus, the apparently larger errors in the parameters for the antipodal solution do not mean that the algorithm failed to find the true peak of the likelihood, but merely that the secondary is shifted from the precisely antipodal position. These results can be directly compared to results obtained using the evolutionary algorithm in~\cite{hea}. The {\sc MultiNest} results are slightly better than those of the evolutionary algorithm, although in both cases the results were within the error we would expect from noise fluctuations and so this difference could easily be due to a difference in the noise realisation used to generate the data. We note that the evolutionary algorithm also recovered parameters for the antipodal solution of the non-coalescing source that were about $4\sigma$ from the true parameters, which is consistent with the prior explanation for that difference. Presently, the  {\sc MultiNest} algorithm runs about five to ten times faster than the evolutionary algorithm, but the latter has not yet been optimized and further work is needed to understand how the efficiency and speed of both algorithms will scale when they are applied to other types of gravitational wave source.

The fact that the intrinsic parameters of the coalescing source are all so well determined is due to a combination of the particular noise realisation used and the strong correlations between the intrinsic parameters. The correlations ensure that if one of the parameters is well-determined then all of the parameters will be well-determined. The results seem surprising because the error is so low in all of the parameters, which, if the parameters were uncorrelated, would require an usually low error in five independent variables. However, in reality we only require an unusually low error in one parameter and then correlations force all the errors to be small. Over many noise realisations we would expect the errors in the intrinsic parameters to wander over the range shown in the posteriors, but this wandering would be correlated between the different intrinsic parameters.

\begin{table}
\begin{center}
\lineup
\item[]\begin{tabular}{@{}*{4}{l}}
\br                              
$$&$t_{c}$/yrs&$\bar{m_{1}}/M_{\odot}$&$\bar{m_{2}}/M_{\odot}$\cr 
\mr
1  & $0.9000 \pm 1 \times 10^{-5}$ & $(2.0005 \pm 0.0016) \times 10^{7}$ & $(1.9995 \pm 0.0014) \times 10^{6}$\cr
1A & $0.9000 \pm 1 \times 10^{-5}$ & $(2.0015 \pm 0.0014) \times 10^{7}$ & $(1.9985 \pm 0.0012) \times 10^{6}$\cr
2  & $1.0200 \pm 2 \times 10^{-4}$ & $(7.2743 \pm 0.0740) \times 10^{6}$ & $(1.7852 \pm 0.0147) \times 10^{6}$\cr
2A & $1.0200 \pm 2 \times 10^{-4}$ & $(7.5472 \pm 0.0727) \times 10^{6}$ & $(1.7329 \pm 0.0134) \times 10^{6}$\cr
\br
\br                              
$$&$\theta$&$\phi$&\cr 
\mr
1&$0.6269 \pm 0.0047$ & $4.7312 \pm 0.0092$&\cr
1A& $2.5131 \pm 0.0048$ & $1.5892 \pm 0.0087$&\cr
2&$2.1267 \pm 0.0060$ & $3.9480 \pm 0.0076$&\cr
2A&$1.0156 \pm 0.0059$ & $0.8076 \pm 0.0077$&\cr
\br

\end{tabular}
\caption{Parameter inferences derived from the first, $\cal{F}$-statistic, stage of the search. The quoted uncertainties are the one sigma errors from the posterior recovered by {\sc MultiNest}. We list the parameter
inferences for both the true and antipodal (`A') sky solutions for each source.}
\label{tab:posterior_stage1}
\end{center}
\end{table}

\begin{table}
\begin{center}
\lineup
\item[]\begin{tabular}{@{}*{7}{l}}
\br                              
$$&&$\bar{M_c}/M_{\odot}$&$\bar{\mu}/M_{\odot}$&$t_{c}/{\rm yrs}$&$\theta$&$\phi$\cr 
\mr
1&$\sigma_{\rm FIM}$&$1.289\times10^3$ &$4.719\times10^3$&$1.209\times10^{-5}$&$6.315\times10^{-3}$&$1.159\times10^{-2}$ \cr
&$\Delta\lambda$&$0.1164$&$0.0763$&$0.0511$&$0.0019$&$0.2036$\cr
1A&$\sigma_{\rm FIM}$&$1.198\times10^3$&$4.405\times10^3$&$1.128\times10^{-5}$&$9.683\times10^{-3}$&$8.529\times10^{-3}$  \cr 
&$\Delta\lambda$&$0.1336$&$0.0795$&$0.9247$&$0.1532$&$0.6597$\cr
2&$\sigma_{\rm FIM}$&$6.986\times10^2$ &$8.642\times10^3$&$3.292\times10^{-5}$&$6.283\times10^{-3}$ &$7.854\times10^{-3}$\cr
&$\Delta\lambda$&$0.7587$&$1.198$&$1.1562$&$0.9742$&$1.0675$\cr
2A&$\sigma_{\rm FIM}$&$7.025\times10^2$&$8.683\times10^3$&$3.301\times10^{-5}$&$6.446\times10^{-3}$&$7.631\times10^{-3}$  \cr 
&$\Delta\lambda$&$3.3452$&$4.0735$&$2.7418$&$0.3818$&$1.0907$\cr
\br
\end{tabular}
\caption{Errors in the intrinsic parameter estimation from the first stage of the search. We quote errors for both the true and antipodal (`A') sky solutions for each source. The first row for each solution gives the one sigma estimation from the Fisher matrix,  $\sigma_{\rm FIM}$., while the second row gives $\Delta\lambda = |(\lambda_{T}-\lambda_{R})/\sigma_{\rm FIM}|$, where $\lambda_T$ denotes the true parameter value, and $\lambda_R$ denotes the recovered value of the parameter.}
\label{tab:mnerrs}
\end{center}
\end{table}

For the second stage of the search, we ran {\sc MultiNest} on the full nine-dimensional parameter space. We ran a separate analysis, each with $500$ `live' points, in the vicinity of each of the modes identified
during the first stage of the search, taking priors on the intrinsic parameters that were $\lambda \in[\tilde{\lambda} - 5\sigma, \tilde{\lambda} + 5\sigma]$, where $\tilde{\lambda}$ and $\sigma$ were the
mean and standard deviation in that parameter as estimated from the posterior recovered in the first stage of the search. In addition, we took natural priors on the four extrinsic parameters --- $\iota \in
[0,\pi]$, $\phi_c \in [0,\pi]$, $\psi \in [0,\pi]$ and $D_L \in [2,15]$Gpc. If the population of supermassive black hole mergers was expected to be uniformly distributed in space, then the natural prior on $D_L$ would be proportional to $D_L^2$. However, in practice the distribution of events will trace the hierarchical assembly of structure and will therefore be complicated and very model-dependent. For that reason, we preferred to use a minimal prior, i.e., a flat prior. This is consistent with the approach used for the MLDC, in which the source SNRs are drawn from a flat prior~\cite{mldc}. The prior on $\phi_c$ is $[0,\pi]$ rather than $[0,2\pi]$ since there is an angle degeneracy in the extrinsic subspace corresponding to the shift $\psi \rightarrow \psi+\pi/2$, $\phi_c \rightarrow \phi_c+\pi$.

\begin{figure}[!tp]
\psfrag{theta}[c][c][0.5]{$\theta$}
\psfrag{phi}[c][c][0.5]{$\phi$}
\psfrag{Mp}[c][c][0.5]{$\bar{m_{1}}$ ($M_{\odot}$)}
\psfrag{Ms}[c][c][0.5]{$\bar{m_{2}}$ ($M_{\odot}$)}
\psfrag{i}[c][c][0.5]{$\iota$}
\psfrag{DL}[c][c][0.5]{$D_{L}$ (Gpc)}
\psfrag{tc}[c][c][0.5]{$t_{c}$ (yrs)}
\psfrag{polarization}[c][c][0.5]{$\psi$}
\psfrag{phase}[c][c][0.5]{$\varphi_c$}
\includegraphics[width=1.2\columnwidth, height=1.2\columnwidth]{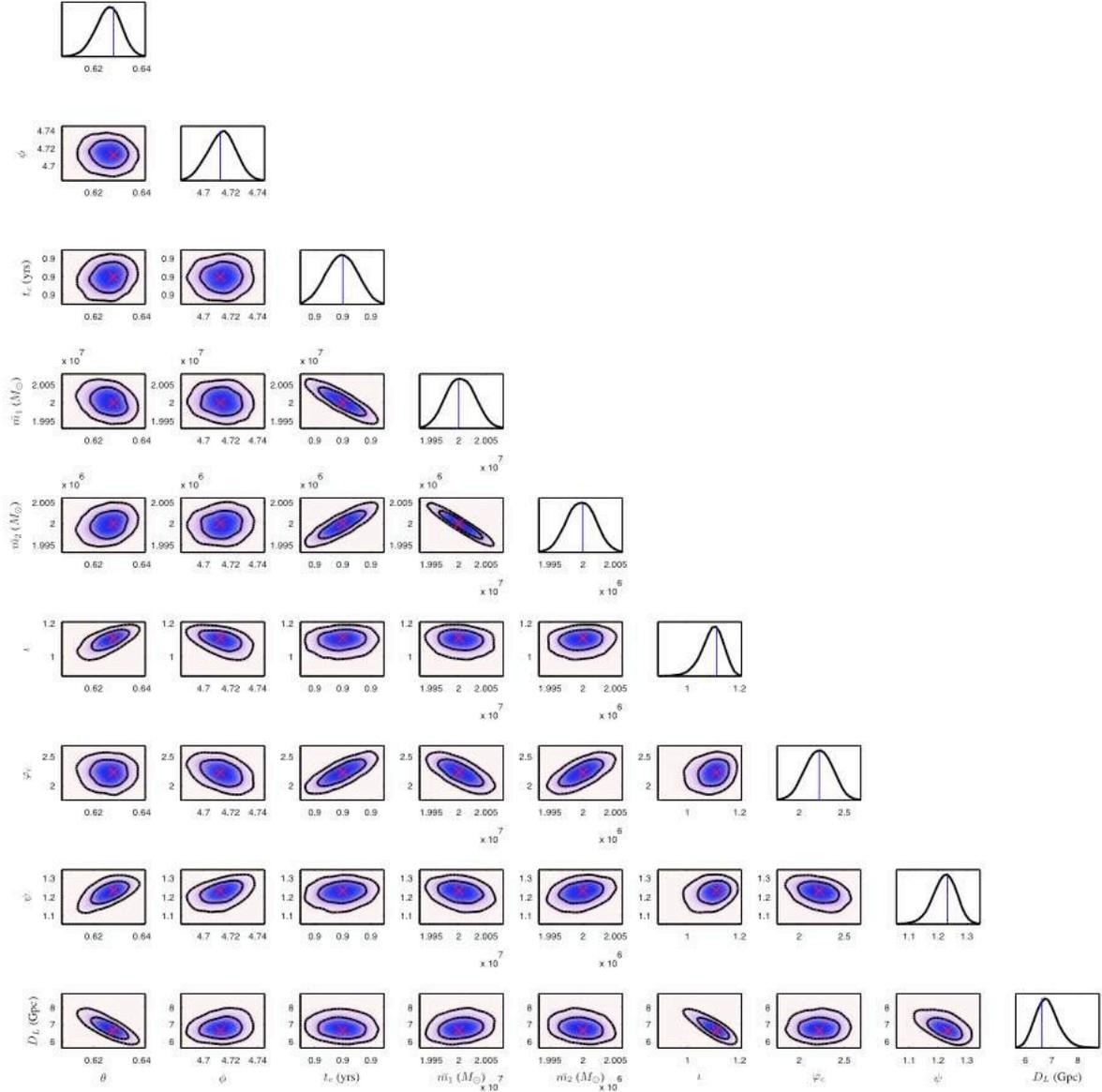}
\caption{Marginalized 1D and 2D posteriors for the primary mode of the bright, coalescing source (source 1)
determined in the second stage analysis using the full likelihood with priors on the intrinsic parameters derived
from the posteriors recovered in the first, $\cal{F}$-statistic, stage. Each plot shows the 2D posterior over the
row (vertical axis) and column (horizontal axis) parameters. At the top of each column is the 1D posterior for
that column parameter. The parameters and the parameter ranges shown in the plots, from top to bottom or left to
right, are $0.63 < \theta < 0.64$, $4.69 < \phi < 4.74$, $0.899992 < t_c < 0.900008$, $1.993\times 10^7 M_{\odot}
< \bar{m_1} < 2.007\times 10^7M_{\odot}$, $1.993\times 10^6 M_{\odot} < \bar{m_2} < 2.007\times 10^6M_{\odot}$,
$0.9 < \iota < 1.2$, $2< \phi_c < 6$ and $1 < \psi < 3$ respectively. The true parameter values are shown by the
blue vertical lines and red crosses overlayed on the 1D and 2D marginalized posterior plots respectively.}
\label{fig:Edfull3}
\end{figure}

\begin{figure}[!tp]
\psfrag{theta}[c][c][0.5]{$\theta$}
\psfrag{phi}[c][c][0.5]{$\phi$}
\psfrag{Mp}[c][c][0.5]{$\bar{m_{1}}$ ($M_{\odot}$)}
\psfrag{Ms}[c][c][0.5]{$\bar{m_{2}}$ ($M_{\odot}$)}
\psfrag{i}[c][c][0.5]{$\iota$}
\psfrag{DL}[c][c][0.5]{$D_{L}$ (Gpc)}
\psfrag{tc}[c][c][0.5]{$t_{c}$ (yrs)}
\psfrag{polarization}[c][c][0.5]{$\psi$}
\psfrag{phase}[c][c][0.5]{$\varphi_c$}
\includegraphics[width=1.2\columnwidth, height=1.2\columnwidth]{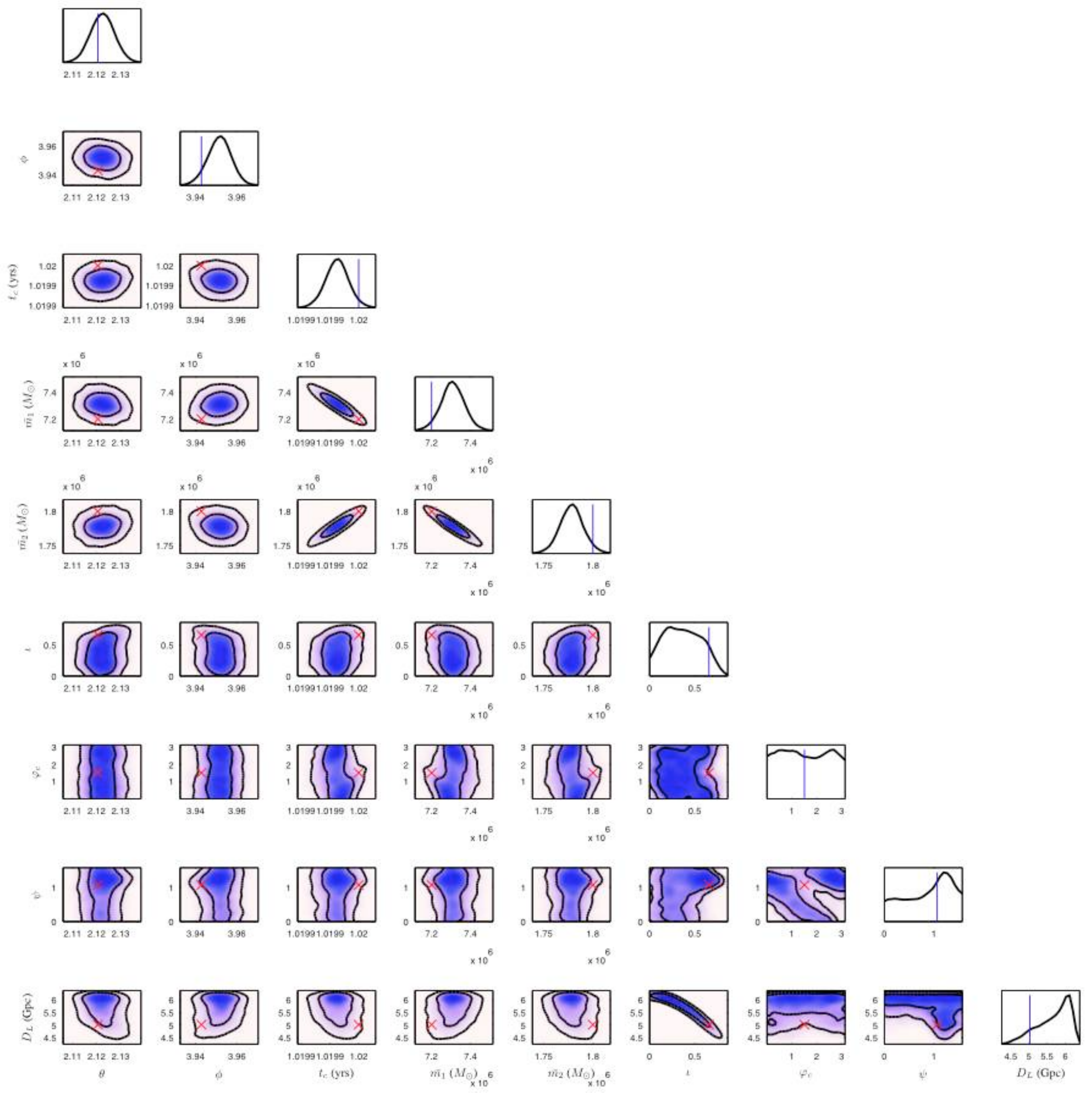}
\caption{As Figure~\ref{fig:Edfull3}, but now for the faint, non-coalescing source (source 2). The parameter ranges shown in the plots, from top to bottom or left to right, are $2.11 < \theta < 2.14$, $3.93 < \phi < 3.97$, $1.01988 < t_c < 1.02004$, $7.1\times 10^6 M_{\odot} < \bar{m_1} < 7.5\times 10^6M_{\odot}$, $1.74\times 10^6 M_{\odot} < \bar{m_2} < 1.81\times 10^6M_{\odot}$, $0 < \iota < 0.8$, $0< \phi_c < 2\pi$ and $0 < \psi < \pi$ respectively. }
\label{fig:Edfull8}
\end{figure}

In the second stage of the search, only one mode of the likelihood was found within the tight priors on the intrinsic parameters provided by the first stage. The second stage of the
search required $\sim 50,000$ likelihood evaluations and took $\sim 30$ minutes on a single 3.0 GHz Intel Woodcrest processor for each mode. The marginalized posterior distributions for the true sky locations of the bright and faint sources are shown in Figs.~\ref{fig:Edfull3} and ~\ref{fig:Edfull8} respectively. 
The posteriors for the faint source are very broad in the extrinsic parameters. This is to be expected, as we do not observe the coalesence for this source and so the error in the estimate of the phase at coalesence, $\phi_c$, is very large, and propagates into other parameters due to correlations. In Table~\ref{tab:mnextres} we quote the values of the extrinsic parameters recovered by the search for the two true modes and the two antipodal modes, and also quote errors as multiples of the one sigma Fisher Matrix error estimate as before. The antipodal solutions have different values for $\iota$ and $\psi$ as well, which are obtained by the transformation $\iota \rightarrow \pi-\iota$, $\psi \rightarrow \pi-\psi$~\cite{arnaud07}, so the true values of the extrinsic parameters at the antipodal sky location are $(\iota, \psi, \theta, \phi) = (2.0296, 1.9086, 2.5133, 1.5708)$ for the coalescing source,``1'', and $(2.4851, 0.5062, 1.0210, 0.8013)$ for the non-coalescing source, ``2''. We do not include the intrinsic parameters in this table as their values did not change significantly in the second stage. The posteriors shown in Figs.~\ref{fig:Edfull3}--\ref{fig:Edfull8} were computed during the second search stage for all parameters.
We see that the algorithm recovers the extrinsic parameters for both sources and both sky solutions to the same high precision as the intrinsic parameters. The posteriors are highly non-Gaussian for the extrinsic parameters of the non-coalescing source in particular, so the Fisher matrix error should not be trustworthy. However, it appears to be giving predictions that are largely consistent with the recovered posteriors.

The {\sc MultiNest} algorithm also returns the evidence associated with each of the recovered modes. These log-evidences were 19,948.5 and 19,948.2 for the true and antipodal modes of the coalescing source and 8,551.3 and 8,555.1 for the true and antipodal modes of the non-coalescing source. These evidence values do not give very strong reason to favour one of the two sky-position modes over the other, but the sky position degeneracy is nearly perfect for the sources we are considering and therefore we would not necessarily expect to be able to distinguish them. However, the evidences of the 7 other modes identified by {\sc MultiNest} were several hundred lower than the evidences of the true and antipodal modes of the source to which they corresponded. There would therefore be good reason to favour the true and antipodal modes over the other modes in the likelihood. If we were using full TDI waveforms at higher frequency we would expect to be able to distinguish between the true and antipodal solutions, and this should be reflected in the evidence values. More investigation is needed to fully understand the utility of evidence for characterisation of modes in the likelihood for these and other gravitational wave sources.

\begin{table}
\begin{center}
\lineup
\item[]\begin{tabular}{@{}*{6}{l}}
\br                              
&&$\iota$&$\psi$&$\phi_c$&$D_L$/Gpc\cr 
\mr
1&$\lambda_R$&$1.1245$&$1.2271$&$2.2203$&$6.5655$ \cr
&$\sigma_{\rm FIM}$&$0.0468$&$0.0442$&$0.5657$&$3.3787$\cr
&$\Delta\lambda$&$0.2668$&$0.1340$&$0.0029$&$0.0203$\cr
1A&$\lambda_R$&$2.0251$&$1.9055$&$2.2119$&$6.5973$\cr
&$\sigma_{\rm FIM}$&$0.0468$&$0.0442$&$0.5657$&$3.3787$\cr
&$\Delta\lambda$&$0.0948$&$0.0694$&$0.0109$&$0.0179$\cr
2&$\lambda_R$&$0.7052$&$2.8670$&$4.7653$&$2.7038$\cr
&$\sigma_{\rm FIM}$&$0.1573$&$0.2201$&$1.4276$&$2.6170$\cr
&$\Delta\lambda$&$0.3096$&$1.0519$&$0.0785$&$0.8866$\cr
2A&$\lambda_R$&$2.4844$&$0.3116$&$4.9330$&$5.7121$\cr
&$\sigma_{\rm FIM}$&$0.1573$&$0.2201$&$1.4276$&$2.6170$\cr
&$\Delta\lambda$&$0.0046$&$0.8839$&$0.2629$&$0.1960$\cr
\br
\end{tabular}
\caption{Maximum a posteriori values of the extrinsic parameters found in the second stage of the search. We quote results for both the two true and the two antipodal sky solutions, as for the intrinsic parameters. For each solution, the first row indicates the recovered value of the parameter, $\lambda_R$, the second row lists the one sigma error estimate from the Fisher Matrix, $\sigma_{\rm FIM}$, and the third row gives $\Delta\lambda = |(\lambda_{T}-\lambda_{R})/\sigma_{\rm FIM}|$, where $\lambda_T$ is the true value of the parameter.} 
\label{tab:mnextres}
\end{center}
\end{table}

\section{Discussion}
\label{sec:discuss}
We have described the use of a multi-modal nested sampling algorithm, {\sc MultiNest}, as a tool for gravitational wave data analysis, illustrating the algorithm by searching for the signals from non-spinning
supermassive black hole binaries in simulated LISA data. We used {\sc MultiNest} to search a data stream containing two such mergers, and the algorithm was able to simultaneously and successfully identify both the true
sky position of each source, and the antipodal sky solution, plus several other secondary modes of the likelihood surface. In a first stage search using the $\cal{F}$-statistic, the algorithm is able to recover the
intrinsic source parameters to very high precision, within $2\sigma$ of the true answer as measured by the theoretical noise-induced error computed from the Fisher Matrix. Following up with a search on the full physical, nine-dimensional, parameter space, the search is also able to recover the extrinsic parameters to similar precision. In addition, the algorithm naturally recovers the posterior distributions and the Bayesian evidence associated with each mode of the solution.

The algorithm works extremely quickly, taking about two and a half hours on a single CPU in this case to complete the whole search. {\sc MultiNest} is fully parallel, which reduces the run time for the whole search to $\sim 20$
minutes when using $10$ processors. More importantly, it achieves these results without using any specific knowledge of the signals for which it is searching --- the waveform model enters only in the likelihood evaluation and not in determining how the `live' point set is updated. While many effective algorithms for searching for non-spinning SMBH binaries are already known, these sources have a multi-modal likelihood which provides a good test case with which to scope out the effectiveness of this algorithm for gravitational wave data analysis. The real power of this algorithm is in its ability to simultaneously identify and characterize all modes of a multi-modal likelihood surface. For signals from EMRIs, the presence of many secondaries in the likelihood has caused problems in data analysis~\cite{EtfAG,neilemri,BBGPa,BBGPb}, and this is likely to be true in searches for spinning black hole binaries as well. {\sc MultiNest} is perfectly adapted to tackling these types of problem, and so these are the next challenges for the algorithm. Research is needed to investigate how the run time, number of
`live' points needed etc. will scale for these other problems. However, the fact that the algorithm performed so well in the non-spinning black hole case, without using any knowledge of the properties of the waveforms, is reason to expect {\sc MultiNest} will also perform strongly in other gravitational wave data analysis applications, for both space-based and ground-based detectors. 

A potentially even more important application of the algorithm in the context of gravitational-wave data analysis will be for evidence evaluation. {\sc MultiNest} computes evidence values as it progresses, which can be used for model selection, e.g., to determine the number of a particular type of source that are present in a data stream or to test alternative theories of gravity etc. Model selection questions have been explored in a LISA context using other algorithms, including MCMC~\cite{littenberg09} and nested sampling~\cite{Veitch:2008wd}. The speed and efficiency of {\sc MultiNest} in answering similar questions should be explored in the future, in order to compare and contrast to these other techniques. The results described in this paper suggest that {\sc MultiNest} will also perform extremely well in such a model evaluation context.
\\

\section*{Acknowledgements}

This work was performed using the Darwin Supercomputer of the University of Cambridge High Performance Computing Service ({\tt http://www.hpc.cam.ac.uk/}), provided by Dell Inc. using Strategic Research
Infrastructure Funding from the Higher Education Funding Council for England and the authors would like to thank Dr. Stuart Rankin for computational assistance. The SMBH waveform and F-Statistic codes used in this work were jointly developed by EKP and Neil J. Cornish at Montana State University. FF is supported by the Cambridge Commonwealth Trust, Isaac Newton and the Pakistan Higher Education Commission Fellowships. JG's work is supported by the Royal Society.

\end{document}